\input{aipcheck}

\documentclass[
    ,final            
  ]
  {aipproc}

\layoutstyle{8x11single}
\usepackage{amsmath}
\usepackage{smartref}

\begin{document}

\title{Long-range correlations in proton-nucleus collisions in MC model with string fusion}


\author{V. Kovalenko}{
  address={Saint Petersburg State University, Russia}
}



\begin{abstract}
The study of long-range correlations between observables in two rapidity windows was proposed as a signature of the string fusion and percolation phenomenon. In the present work we calculate the correlation functions and coefficients for $p\text{--}Pb$ collisions in the framework of the Monte Carlo string-parton model, based on the picture of elementary collisions of color dipoles. It describes pA and AA scattering without referring to the Glauber picture of independent nucleons collisions and includes effects of string fusion. Different types of correlations are considered: $n\text{--}n$, $p_t\text{--}n$, $p_t\text{--}p_t$, where $n$ is the event multiplicity of charged particles in a given rapidity window and $p_t$ is their mean transverse momentum. The results, obtained in the present work for $p\text{--}Pb$ collisions, are compared with available experimental results for $p_t\text{--}n$ correlations in one window and further predictions are made.
\end{abstract}

\maketitle


\section{Introduction}


The present work is devoted to study of long-range correlations in
proton-lead collisions at the LHC energy within the Monte Carlo model. 
The existence of long-range correlations (also referred as forward-backward
correlations) between observables in separated rapidity intervals
was established, and they has been measured in a wide range of energies and
different colliding systems (including pp and AA collisions) (see refs in \cite{PoSFeofilov}). The presence of the correlations between observables in
rapidity intervals separated by a wide rapidity gap is attributed to
the concept of quark-gluon strings (or flux tubes) \cite{F31}, that might be formed
 at the very early stages of a hadron-hadron collision. The strings
 conception, which comes from the Regge-Gribov approach \cite{WERNER},
 implies that extended objects are stretched between 
 interacting constituents and decay,
 contributing the charged particles in a wide rapidity range (which
 at the LHC energies might be as large as 10 rapidity units).
The major physics motivation for the detailed study of long-range correlations \cite{F1}
is a possibility of the interactions between several strings \cite{F32}, producing
new sources with higher tension \cite{F33}. This mechanism, called string fusion, 
could be one of the approaches leading to the formation of quark-gluon plasma \cite{OL1}.

It was proposed \cite{VK1, VK2} to study three types of correlations:
\vspace{0.15cm}

\hspace{0.3cm}(i) $n\text{--}n$ -- correlations between number of charged particles in two rapidity intervals;

\hspace{0.3cm}(ii) $p_t\text{--}n$ -- correlation between charged multiplicity in one window and mean transverse momentum in another window;

\hspace{0.3cm}(iii) $p_t\text{--}p_t$ -- correlation between mean transverse momentum
of charged particles in two windows.
\vspace{0.15cm}

One should stress, that multiplicity-multiplicity correlations are present
also in models without string fusion, due to event-by-event fluctuations \cite{VecherninPoS,VecherninAIP}
in the number of (identical) strings, while for non-zero
 $p_t\text{--}n$ and $p_t\text{--}p_t$ correlations a mixture
 of the sources of different types (ex. strings of different tension)
 is essential.

In order to characterize long-range correlations
numerically, a concept of correlation function and
correlation coefficient was introduced 
\cite{Vest1,Vest2,OL9}. A correlation function is defined as
the mean value $\langle B \rangle_F$ of variable B
in the backward window as a function of
another variable F in the forward rapidity
window:

\begin{equation}
f_{B\text{--}F}(F)={\langle B\rangle}_F.
\end{equation}

The correlation coefficient is a slope
of correlation function:
\begin{equation} \label{defcorr}
b_{B\text{--}F}=\frac{df_{B\text{--}F}(F)}{dF}|_{{F}=<{F}>}.
\end{equation}

It is useful often  to switch to
normalized variables: $B \rightarrow B/{\langle B\rangle},$ $F \rightarrow F/{\langle F\rangle}$, in this case
 $p_t\text{--}n$ correlation coefficient  becomes dimensionless,
 and both $n\text{--}n$ and $p_t\text{--}p_t$ correlation coefficients
do not change in case of symmetrical windows.
Thus we use the following definitions:
\begin{eqnarray}
\label{normnn} b_{n\text{--}n}^{}&=&\frac{<n_F>}{<n_B>} \cdot\frac{d<{n}_B>}{dn_F}|_{n_{F}=<n_{F}>}, \\
\label{normptn} b_{p_t\text{--}n}^{}&=&\frac{<{n}_F>}{<{p_t}_B>} \cdot\frac{d<{p_t}_B>} {d{n}_F}|_{n_{F}=<n_{F}>}, \\
\label{normptpt} b_{p_t\text{--}p_t}^{}&=&\frac{<{p_t}_F>}{<{p_t}_B>} \cdot\frac{d<{{p_t}}_B>}{d{p_t}_F}|_{{p_t}_{F}=<{p_t}_{F}>}.
\end{eqnarray}

In practice, determination of the correlation coefficient in Monte Carlo simulations is performed by obtaining the fit of $f(F)$ by linear function from ${\langle F\rangle}-\sigma_{{\langle F\rangle}}$ to ${\langle F\rangle}+\sigma_{{\langle F\rangle}}$,
where $\sigma_{{\langle F\rangle}}=\sqrt{\langle F^2 \rangle-{\langle F\rangle}^2}$. This method is used in this paper.

For the purposes of quantitative theoretical study of 
 the effects of string fusion on 
observables, a parton-string Monte Carlo model
has been developed for high-energy proton-proton and heavy ion
collisions \cite{OL2,OL3}. The model describes soft part
of multiparticle production in a wide energy range
without referring to the Glauber picture of independent nucleons collisions.
The model was applied to $pp$ and $Pb\text{--}Pb$ collisions at the LHC energy \cite{OL3,OL15},
and predictions on correlation coefficients have been made.
The present work focuses on $p\text{--}Pb$ collisions at 5.02 TeV.

The paper is organized as follows: in the next section we give a very brief description of the Monte Carlo model with string fusion. After that we present and discuss the results obtained, and also compare them to available experimental data for $p_t\text{--}n$ correlation
in one window. Finally, we present our conclusions.

\section{Non-Glauber Monte-Carlo model \cite{OL2, OL3}}

The present model describes nucleon-nucleon collisions on the partonic level.
It is supposed, that initially the nucleons are distributed in $Pb$ nuclei according to Woods-Saxon distribution:
$$\rho(r)=\frac{\rho_0}{1+\exp[(r-R)/d]},$$
with parameters $R=6.63$ fm, $d=0.545$ fm \cite{WoSa}.

Each nucleon supposed to consist of a valence quark-diquark pair and certain number of sea quark-antiquark pairs, distributed around the center of nucleon according to two-dimensional Gauss distribution with mean-square radius $r_0$.  The number of sea pairs is generated according to Poisson distribution.

An elementary interaction is realized in the model of colour dipoles \cite{OL7, OL8}.
  It is assumed that quark-diquark and quark-antiquark pairs form a dipoles.
The probability amplitude of the collision of two dipoles from target and projectile is given by:
\begin{equation} \label{newformula}
			f=\frac{\alpha_S^2}{2}\Big[ K_0\left(\frac{|\vec{r}_1-\vec{r}_1'|}{r_{\text{max}}}\right) +
			K_0\left(\frac{|\vec{r}_2-\vec{r}_2'|}{r_{\text{max}}}\right) 
			- K_0\left(\frac{|\vec{r}_1-\vec{r}_2'|}{r_{\text{max}}}\right)
			- K_0\left(\frac{|\vec{r}_2-\vec{r}_1'|}{r_{\text{max}}}\right)	\Big]^2.
\end{equation}	
where $K_0$ is modified Bessel function of the second order, $(\vec{r}_1, \vec{r}_2), (\vec{r}_1', \vec{r}_2')$ are transverse coordinates of the projectile and target dipoles, and $\alpha_s$ --  effective coupling constant,
$r_{\text{max}}$ is characteristic scale, responsible for the confinement effects.
One can observe that two dipoles interact more probably, if the ends are close to each other, and (others equal) if they are wide.

In our Monte Carlo model it is assumed, that if there is a collision between two dipoles,
 two quark-gluon strings
are stretched between the ends of the dipoles,
and the process of string fragmentation gives
observable particles. The particle production of a string is assumed to go  uniformly between the string rapidity ends $y_{\text{min}}$ and $y_{\text{max}}$ \cite{OL2,OL6}, with mean number of charged particles per rapidity $\mu_0$, and independently in each rapidity interval, with Poisson distribution.
Note that in the present model every parton can interact with other one only once, forming a pair of quark-gluon strings, hence, producing particles, which contradicts to Glauber supposition of constant nucleon cross section and accounts of energy conservation
in the process of the nucleons collision \cite{OL14,QftPA}.

The transverse position of string is assigned to the arithmetic mean
of the transverse coordinates of the partons at the ends of the string.
Due to finite transverse size of the strings they overlap, that in the framework of string fusion model
 \cite{OL9, OL10} gives a source with higher tension.
We used  the cellular variant of the model \cite{VK1, VK2, Vest1, Vest2, VecherninPoSLak}, according to which a lattice in the transverse plane
is introduced, with the area of a cell being equal to the transverse string
area ($S_{\text{str}}=\pi r^2_{\text{str}}$), and strings are
supposed to be fused if their centres occupy the same cell. Mean multiplicity of charged particles and mean $p_t$ originated from the cell where $k$ strings are overlapping are the following:
\begin{equation} \label{muptloc}\nonumber
	\left\langle \mu\right\rangle_k=\mu_0 \sqrt{k}, \hspace*{1cm}
	\left\langle p_t\right\rangle_k=p_0 \sqrt[4]{k}.
\end{equation}	 

Here $\mu_0$ and $p_0$ are mean charged multiplicity from one single string per rapidity unit
and mean transverse momentum from one single string. 
This relations allows to calculate long-range correlation functions and
correlation coefficients between
multiplicities and mean transverse momentum of charged particles \cite{OL3}.
Important, that while using of the normalized variables (\ref{normnn}) -- (\ref{normptpt})
these parameters  ($\mu_0$,$p_0$) cancels out and do not influence the final result,
which makes the calculations more robust. 

Parameters of the model in general are constrained \cite{OL6} from the pp data on the total inelastic cross section and charged multiplicity in wide energy range (from ISR to LHC) with additional requirement of  consistent description of the multiplicity in minimum bias $p\text{--}Pb$ and $Pb\text{--}Pb$ collisions at the LHC energy.
The following set of parameters is used in the present work \cite{OL6}:
\begin{center}
\begin{tabular}{lcccc}
\hline
    {$r_\text{str}\mathsf{, fm}$ }
  & {$r_0\mathsf{, fm}$}
  & {$\alpha_s$}
  & {$r_\text{max}/r_\text{0}$}
  & {$\mu_0$}   \\
\hline
0   & 0.6 & 0.4 & 0.9 & 1.010 \\
0.2 & 0.6 & 0.5 & 0.4 & 1.152 \\
0.3 & 0.6 & 0.6 & 0.2 & 1.308 \\
\hline
\end{tabular}
\end{center}

 Centrality in the present work is determined as a fraction of the events with observable among the whole distribution. For the centrality estimator a multiplicity signal in 
 forward rapidity region  is used:  a sum of charged multiplicities in rapidity windows: (3.0; 5.0)+(-3.6; -1.6), which is in correspondence to the coverage of the ALICE detector VZERO \cite{OL12},  used as centrality estimator in the ALICE experiment \cite{OL13}.

\begin{figure}[h]\label{Fig1}
  \includegraphics[height=.334\textheight]{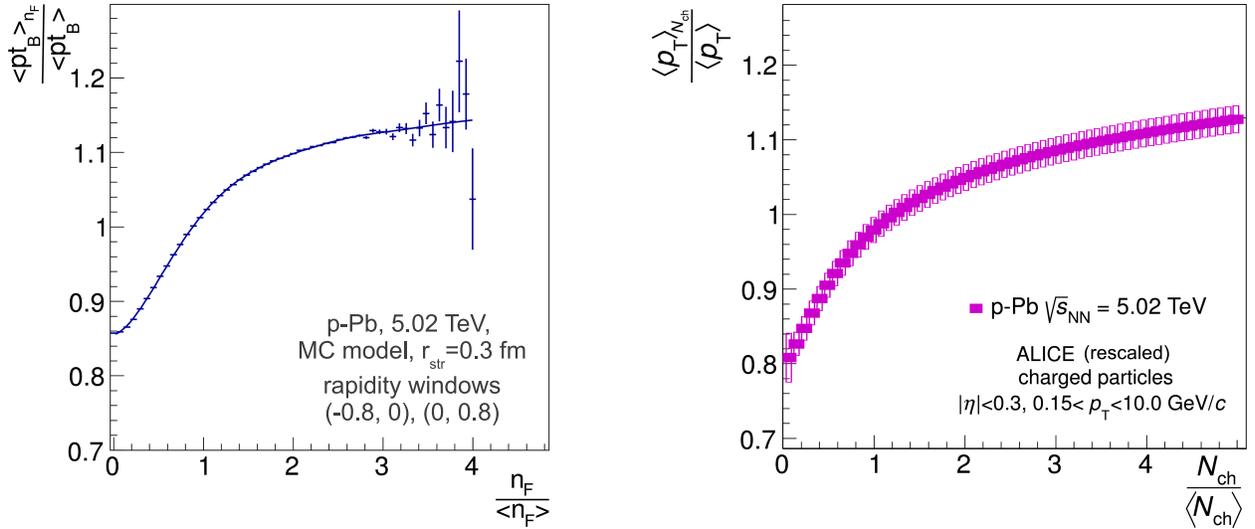}
  \caption{$p_t\text{--}n$ correlation function in $p\text{--}Pb$ collisions at $\sqrt{s}=$5.02 TeV: left -- results of the MC model with sting fusion, for $p_t\text{--}n$ correlation in
  separate rapidity windows in normalized variables; right -- $p_t\text{--}n$ correlation 
  function in one pseudorapidity window, measured by ALICE \cite{OL5} and plotted here
  in relative variables for the comparison.}
\end{figure}

\section{Results}

\subsubsection{ pt-n correlation functions}

At the Fig. \ref{Fig1} (left) the result of the
calculation of $p_t\text{--}n$ correlation 
function in rapidity windows (-0.8, 0), (0, 0.8) is shown. The plot
was obtained for minimum bias $p\text{--}Pb$ collisions in the model
with string fusion and with string radius $r_\text{str}=0.3$ fm.
The normalized variables were used. Considerable
relative growth of the transverse momentum with multiplicity is observed, 
with tendency of saturation (flattening) at high multiplicity.

The right plot of Fig. \ref{Fig1} shows experimental data \cite{OL5} on the
$p_t\text{--}n$ correlation, rescaled to normalized variables.
The comparison of the model results and the data is limited,
due to the restriction on the transverse momentum $0.15<p_t<10 \text{ GeV}/c$
in the ALICE data, and the possible influence of short-range effects,
contributing to the $p_t\text{--}n$ correlation function in one window.
Taking this into account, one observe reasonably good agreement
between the model and experimental data.

From this comparison we can make two conclusions: (i) the string fusion is
a good candidate to the mechanism, responsible to the growth
of the transverse momentum with multiplicity at the LHC energy;
(ii) the long-range correlations seem to dominate over short-range effects even
at zero rapidity gap between windows (in fact, the present MC model
do not contain any short-range effects). Note that the latter
supports similar conclusion, obtained while analysing
the LHC data on multiplicity-multiplicity forward-backward correlations \cite{VecherninAIP}.

\begin{figure}[h]\label{Fig2}
  \includegraphics[height=.334\textheight]{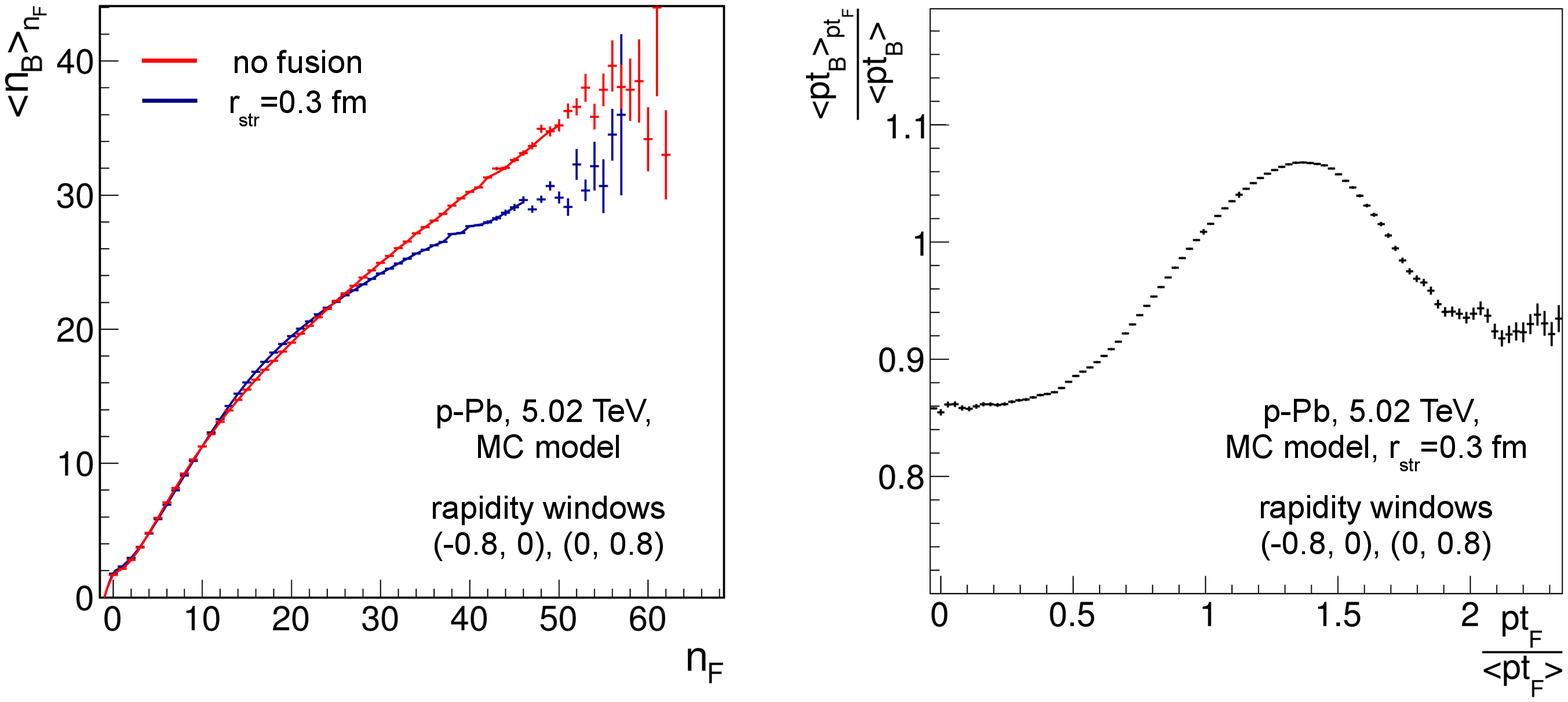}
  \caption{$n\text{--}n$ (left) and $p_t\text{--}p_t$ (right) correlation functions minimum bias $p\text{--}Pb$ collisions at $\sqrt{s}=$5.02 TeV, calculated in the MC model.}
\end{figure}

\subsubsection{ n-n,  pt-pt correlation functions}

At the Fig. \ref{Fig2} the calculated
$n\text{--}n$ and $p_t\text{--}p_t$ correlation functions are shown.
The rapidity intervals are the same, as for $p_t\text{--}n$ correlation function;
$n\text{--}n$ correlation function is plotted in the absolute
variables, while $p_t\text{--}p_t$  -- in relative.

Multiplicity-multiplicity correlation function is found 
to be non-linear both with and without string fusion.
The deviation from the linearity and saturation of n-n correlation
 at high multiplicities is
more pronounced in the model with fusion, that could be
an indication of the string fusion.

The correlation function between mean transverse momenta
is found non-linear and even non-monotonic in minimum bias 
$p\text{--}Pb$ collisions. At present, there is
no experimental data on these observables,
but the shape of the correlation function
can be good experimental test for the string fusion model.

\begin{figure}[h]\label{Fig3}
  \includegraphics[height=.305\textheight]{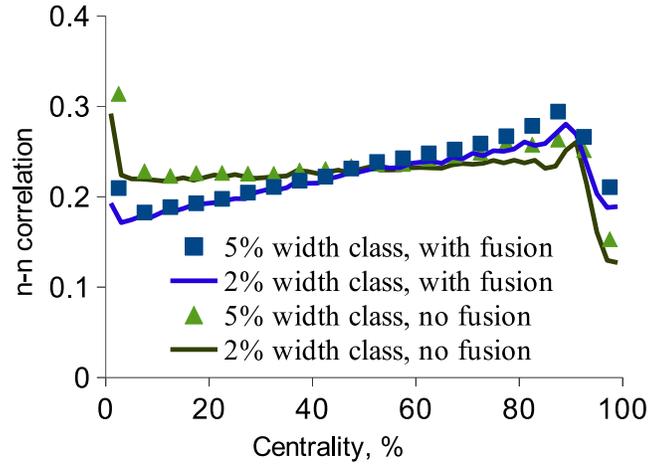}
  \caption{Centrality dependence of $n\text{--}n$ correlation coefficient
  in $p\text{--}Pb$ collisions
  at $\sqrt{s}=$5.02 TeV in the MC
  model with and without string fusion.}
\end{figure}

\begin{figure}[h]\label{Fig4}
  \includegraphics[height=.275\textheight]{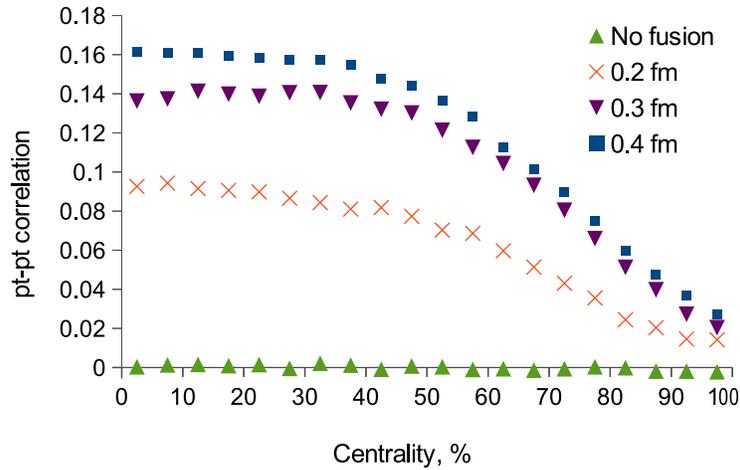}
  \caption{Dependence of $p_t\text{--}p_t$ correlation coefficient
on centrality at different string transverse radius. Calculations in MC model, $p\text{--}Pb$ collisions
  at $\sqrt{s}=$5.02 TeV.}
\end{figure}

\subsubsection{Centrality dependence of n-n correlation coefficient}

Results for multiplicity-multiplicity  correlation coefficient as a function of centrality are shown in Fig. \ref{Fig3}. It is found that $n\text{--}n$ correlation
coefficient decreases with centrality. We
note, that similar behaviour of 
$n\text{--}n$ correlation coefficient
is present AA collisions \cite{OL15}
at the LHC energy in the model with string fusion.
Without the string fusion,
the correlation coefficient is
nearly constant with centrality, that indicates that the decrease of $b_{n\text{--}n}$ is actually an effect of string fusion. The variation of the width of centrality class from 2\% to 5\% do
not influence the value of the correlation
coefficient.

%

\subsubsection{Dependence of pt-pt correlations
on the  transverse radius of string}

The value of the $p_t\text{--}p_t$ correlation coefficient as a function
of centrality is shown in the Fig. \ref{Fig4}. 
Flat behaviour in central and semi-central collisions is observed, with decrease in
the peripheral part. We note that the results are sensitive to the transverse radius of string: the thicker the
strings, the more probable their
overlapping and fusion, that
gives higher correlation coefficient.
Comparison of this predictions on $p_t\text{--}p_t$ correlations with experimental data would give
a good opportunity to constrain the
value of string
transverse radius.


\section{Conclusions}

 The Monte Carlo  model with string fusion is applied to the
  calculation of long-range correlations 
  in $p\text{--}Pb$ collisions at $\sqrt{s}=$5.02 TeV.
  Both correlation functions and correlation coefficients were 
  studied.
Thus it was possible to establish:

\hspace{0.3cm}(i) $P_t\text{--}n$ correlation function is in qualitative agreement with the
experimental data in one window.

\hspace{0.3cm}(ii) Multiplicity correlation function tends to saturate in the model
with string fusion.

\hspace{0.3cm}(iii) $b_{n\text{--}n}$ considerably decreases with centrality in case of string fusion
and remain flat with centrality in the model without fusion.

\hspace{0.3cm}(iv) $P_t\text{--}p_t$ correlations have a plateau in central and semi-central
collisions and decrease in periphery.

The string fusion is
a good candidate to the mechanism, responsible to the growth
of the transverse momentum with multiplicity at the LHC energy.
The experimental data on the centrality dependence of $pt\text{--}pt$
correlations would be a good test of string fusion model,
allowing to constrain the transverse radius of string.


\begin{theacknowledgments}

 The author is grateful to V.V.Vechernin and G.A.Feofilov for useful discussions and to C.Pajares for his interest to this work.  The support of SPbSU by grant 11.38.66.2012 and Special Rector's Scholarship is acknowledged.

\end{theacknowledgments}

\bibliographystyle{aipproc}   

\end{document}